\documentstyle[12pt]{article} %especifica o estilo do documento 

\title{\bf Digital Spectrometry Signal Treatment Applied to a Fiber Optic Resonant Gyroscope for Rate Measurements}        %Título do documento.
\author{\bf R. T. Almeida \footnote{rtorres@ipd.eb.mil.br}, O. B. Campos \footnote{comar@ipd.eb.mil.br} 
\\ \bf J. M. Magalhães, J. F. M. Mendes, M.C. Carvalho  
\\ \bf W. V. Santos \footnote{swillian@ipd.eb.mil.br}, G. G. Filho \footnote{ggurgel@ipd.eb.mil.br}, R. C. B. Allil \footnote{rallil@ipd.eb.mil.br}
\\
\\
\bf Instituto de Pesquisa e Desenvolvimento- IPD
\\
\bf Centro Tecnol\'ogico do Ex\'ercito - CTEx 
\\
\\
\bf Av.das Am\'ericas, 28.705, Guaratiba
\\
\bf Rio de Janeiro
\\
\bf Brazil 
\bf ZipCode: 23020-470}         % Declara o nome do autor.

\begin{document}           % Fim do preâmbulo e início do texto.

\large 

\maketitle                 % Produz o título.
\vfil 
\pagebreak
\abstract

\hspace{.5cm}The FORG \footnote{\em Fibre Optic Resonant Gyroscope} opeeration principle uses a recirculating ring resonant cavity to get a rotation-induced Sagnac effect enhancement \cite{ezekiel}. It grants to a FORG a comparable sensitivity in relation an I-FOG \footnote{\em Interferometric Fibre Optic Gyroscope}that has the fiber lenght $\Im /2$ times longer, where $\Im$  is equal its finesse factor. Other advantages is despite of thermal drift because the FORG uses less quantity of fibre than the I-FOG, giving to the first less thermal drift sensitivity than the last. But, due to the Kerr effect and couplers loss, that causes parasitics effects like dissimetries on signal response \cite{youngquist} and cross polarization. due to these facts, the FORG complexity is increased a lot. The signal response dissimetry caused by kerr effect can be corrected by phase nulling method, was proposed by Hotate \cite{hotate}. The proposal of this work is to show a applied to a FORG technique that simplifies the signal treatment, employing all digital setup, like as filter banks and wavelets methods, resulting in a maximally flat scale factor. In this investigation are presented over the simulations results, employing the modified digital FM spectrometry techniques  by decimation and interpolation techniques over  a ring resonator that pursuit a 10 meters SM-PM lenght fiber coil and 10 centimeters of diameter, with a $1.55 \mu m$ laser source. The advantages of these techniques are to simplify the electronic circuitry, offering an upgrade facility, using only one DSP (Digital Signal Processor), realizing all needed functions. The investigation of this method is based in a optical field switching scheme and digital frequency domain spectrometry. The purpose of this work is to describe this digital technique, well as the simulation results, discussing about this technique use and its limitations.

\section*{\large \bf 1. Introduction}
\vspace{.5cm}
\hspace{.5cm}The FORG operation principle is based on a recirculating light into the resonant cavity or a fibre ring resonator. It increases its sensitivity, enhancing the rotation induced Sagnac effect \cite{ezekiel} by a $\Im/2$, where $\Im$ is equal its finesse factor. In this work we intend to show one easy solution to signal process employing less hardware. This technique employs light switching and digital frequency decimation methods to get the rate measurement signal.

\section*{\large \bf 2. Theoretical Background}
\vspace{.5cm}
\hspace{.5cm}The proposed setup is depicted in figure 1, that uses two electro-optic modulators, one in each Mach-Zehnder interferometer branch, each of wich working as an amplitude modulator (see ref.\cite{saleh} for more details). Jointly of these modulators, have two electronic switches working syncrinizatelly. The polarization controllers and optical isolators had been supressed of fig.1 for simplicity. The set of equations that describe the FORG theory will be given in the following pharagraphs. The FORG response equations can be found in reference \cite{stokes}, which will be given below. 

\begin{equation}
R_{i}(\beta)=\bigl(1-\gamma_{0}\bigr) \times \Biggl[1-\frac{\bigl(1-\kappa_{r}\bigr)^{2}}{\bigl(1+\kappa_{r}\bigr)^{2}-4\kappa_{r} \sin^{2}\bigl(\frac{\beta_{i}L}{2}-\frac{\pi}{4}\bigr)}\Biggr] \
\label{transf1}
\end{equation}

Or in therm of $\Delta L_{r}$ and $\sigma_{0}$ \cite{lefevre}:

\begin{equation}
P_{i}(\Omega)=\bigl(1-\gamma_{0}\bigr) \times \Biggl[1-\frac{\bigl(1-\kappa_{r}\bigr)^{2}}{\bigl(1+\kappa_{r}\bigr)^{2}-4\kappa_{r} \sin^{2}\bigl(\pi \Delta L'_{R}(\Omega) \sigma_{0} \bigr)}\Biggr] \
\label{transf2}
\end{equation}

Where, in eq.\ref{transf2}, $\Delta L'_{R}(\Omega)=\frac{L_{R}D\Omega}{2c_{0}}$, and $\sigma_{0}=\frac{1}{\lambda_{0}}$. The values $\gamma_{0}$, $\kappa_{0}$ are, respectively, the factional coupler intensity loss and resonant coupling coefficient ($\kappa_{r}=(1-\gamma_{0})^{-2 \alpha_{0} L}$, where $\alpha_{0}$ is the exponential attenuation per lenght unit). The index $i$ in $R_{i}(\beta)$ denotes the $i^{th}$ port ($i=1, 2$) output. These equations are derived from coupling matrix, depicted in equation \ref{transf5}:

\begin{equation}
\frac{E_{ccw}}{E_{0}} = A \{ {\bf C} + A {\bf B}^{2}  \alpha \sum^{\infty}_{m=1}\bigl[ \alpha A {\bf C} \bigr]^{m-1}e^{-j ( - \omega_{0}m \tau \sum^{m}_{k=0} \phi^{-}(t-k \tau )) }  \}
\label{transf3}
\end{equation}
\begin{equation}
\frac{E_{ccw}}{E_{0}} = A \{ {\bf C} + A {\bf B}^{2}  \alpha \sum^{\infty}_{m=1}\bigl[ \alpha A {\bf C} \bigr]^{m-1}e^{-j ( - \omega_{0}m \tau \sum^{m-1}_{k=0} \phi^{+}(t-k \tau )) }  \}
\label{transf4}
\end{equation}

Where the A, B, C and D constants are the complex elements of the unitary coupling matrix, and $\alpha$ is equal the attenuation due to the fiber lenght ($\alpha = \exp [ -\alpha_{0} L ] $). The coupling matrix is shown below

\begin{equation}
\bf{\cal{C}}= (1- \gamma_{0})^{2} \Biggl(  \begin{array}{cc}
                                           \bf A & \bf B \\
			                         \bf C & \bf D
			                         \end{array} \Biggl) 
\label{transf5}
\end{equation}

The {\bf A} and {\bf B} coefficients are equal to $(1- \kappa)^{\frac{1}{2}}$, and {\bf C} and {\bf D} coefficients are equal to $j \kappa^{\frac{1}{2}}$. The $kappa$ constant is the intensity coupling coefficient. Taking account the phase modulator characteristics (e.g.: an $LiNbO_{3}$ electro-optic modulator or a fiber wrapped and bounded in a piezoelectric cylinder), its response has the following equation:

\begin{equation}
\Delta L_m = \frac{\lambda_0 \beta_m \sin(\omega_m t)}{2 \pi}
\label{modulador}
\end{equation}

Changing the term in summation into the exponential in eq. \ref{transf3} and \ref{transf4}, by eq. \ref{modulador}, after some algebraic manipulations, these expressions can be given by (see reference \cite{gradshteyn} for more details):

\begin{eqnarray}
\sum^{m}_{k=1} \phi^{-}_{m} & = & \frac{\beta_m}{2} \frac{\sin \lbrack \omega_m (t+m \tau_d)  \rbrack - \sin \lbrack \omega_m (t-m \tau_d)  \rbrack + \sin \lbrack \omega_m (t- \tau_d)  \rbrack }{1- \cos( \omega_m \tau_d )}
\label{respmod1} \\
\sum^{m-1}_{k=1} \phi^{+}_{m} & = & \frac{\beta_m}{2} \frac{ \sin \lbrack \omega_m (t- \tau_d)  \rbrack - \sin ( \omega_m t) - \sin \lbrack \omega_m (t+ m \tau_d)  \rbrack - \sin \lbrack \omega_m (t+ (m-1) \tau_d)  \rbrack}{1- \cos( \omega_m \tau_d )}
\label{respmod2}
\end{eqnarray}

To get a maximum sensitivity rotation response(and simplifying the computational cost), the equations \ref{respmod1} and  \ref{respmod2} must obeys the following condition:

\begin{equation} 
\omega_m \tau_d = \frac{\pi}{2} \Rightarrow f_m = \frac{c_0}{4n_0 L_r}
\label{freqmod}
\end{equation}

The resulting frequency in eq. \ref{freqmod} is the necessary frequency to reach the maximum phase variation. Expanding the expression \ref{transf2} in Taylor's series, into the discrete-time domain ($t \rightarrow nT_s$), changing the values into the sine, already depicted in eq. \ref{transf1}, taking account the frequency value in eq. \ref{freqmod}, we get the following expressions:

\begin{eqnarray}
P_{ccw}(\Omega) & = & \Gamma^{0}_{k} + \Gamma^{1}_{k}( \pi \Delta L'_{R} ( \Omega ) \sigma_0 ) + \Gamma^{2}_{k}( \pi \Delta L'_{R} ( \Omega ) \sigma_0 )^2 + \Gamma^{3}_{k}( \pi \Delta L'_{R} ( \Omega ) \sigma_0 )^3 + {\cal{O}}(4) 
\label{resposta1} \\
P_{cw}(\Omega) & = & \Gamma^{0}_{k} - \Gamma^{1}_{k}( \pi \Delta L'_{R} ( \Omega ) \sigma_0 ) + \Gamma^{2}_{k}( \pi \Delta L'_{R} ( \Omega ) \sigma_0 )^2 - \Gamma^{3}_{k}( \pi \Delta L'_{R} ( \Omega ) \sigma_0 )^3 + {\cal{O}}(4) 
\label{resposta2}
\end{eqnarray}

Where the $\Gamma^{0}_{k}$, $\Gamma^{1}_{k}$, $\Gamma^{2}_{k}$ and $\Gamma^{3}_{k}$ coefficients are, respectivelly:

\begin{eqnarray}
\Gamma^{0}_{k} & = & \frac{A'}{2} \frac{(2C'-2B'-E')+C'^2 \cos(2 \phi (k))}{C'-E' \sin^2 (\phi (k))} \label{gama0} \\
\Gamma^{1}_{k} & = & \frac{A'B'E' \sin(2 \phi (k))}{(C'-E' \sin^2 (\phi (k)))^2}  \label{gama1} \\
\Gamma^{2}_{k} & = & \frac{A'B'E' \cos(2 \phi (k))}{(C'-E' \sin^2 (\phi (k)))} + \frac{A'B'E'^2 \sin(2 \phi (k))}{(C'-E' \sin^2 (\phi (k)))^2} \label{gama2} \\
\Gamma^{3}_{k} & = & \Biggl[ \frac{2A'B'E' \sin(2 \phi (k))}{(C'-E' \sin^2 (\phi (k)))^2} - \frac{A'B'E'^2 \sin(4 \phi (k))}{(C'-E' \sin^2 (\phi (k)))^3} - \nonumber\\ 
& & \frac{A'B'E'^2}{4} \frac{(2C'-3E') \sin(4 \phi (k))+10E' \sin( \phi (k))}{(C'-E' \sin^2 (\phi (k)))^4} \Biggr] \label{gama3} 
\end{eqnarray}

The $ \phi $ function is equal $ \beta_m sin \lbrack \frac{2 \pi k}{N} \rbrack $, where $\beta_m$  is the modulation index, and $\frac{2 \pi k}{N}$ is the discrete modulation frequency. The A', B', C' and E' Coefficients of \ref{gama0} to \ref{gama3} are described as

\begin{eqnarray}
A' & = & 1- \gamma_0 \\
B' & = & (1- \kappa_0)^2 \\
C' & = & (1+ \kappa_0)^2 \\
E' & = & 4 \kappa_r
\end{eqnarray}

The sine and cosine terms in eqs. \ref{gama0}, \ref{gama1}, \ref{gama2} and \ref{gama3} are expandable in $1^{st}$ kind of Bessel's series. These series, in sine and cosine expansion, are given by:

\begin{eqnarray}
\sin( \phi_k) & = & 2 \sum^{\infty}_{n=1}J_{2n-1} ( \beta_m) \sin \Biggl[ 2k \pi \frac{2n-1}{N} \Biggl] \label{bessel1} \\
               & = & 2 \cos \Biggl( \frac{2k \pi}{N} \Biggr) \sum^{\infty}_{n=1} J_{2n-1} ( \beta_m) \sin \Biggl[ \frac{4nk \pi}{N} \Biggr] - 2 \sin \Biggl( \frac{2k \pi}{N} \Biggr) \sum^{\infty}_{n=1} J_{2n-1}( \beta_m) \cos \Biggl[ \frac{4nk \pi }{N} \Biggr]\label{bessel2} \\
\sin(2 \phi_k) & = & 2 \sum^{\infty}_{n=1} J_{2n-1}( 2\beta_m) \sin \Biggl[ 2k \pi \frac{2n-1}{N} \Biggl] \label{bessel3} \\
              & = & 2 \cos \Biggl( \frac{2k \pi}{N} \Biggr) \sum^{\infty}_{n=1} J_{2n-1}(2 \beta_m) \sin \Biggl[ \frac{4nk \pi}{N} \Biggr] - 2 \sin \Biggl( \frac{2k \pi}{N} \Biggr) \sum^{\infty}_{n=1} J_{2n-1}(2 \beta_m) \cos \Biggl[ \frac{4nk \pi }{N} \Biggr]\label{bessel4} \\
\sin(4 \phi_k) & = & 2 \sum^{\infty}_{n=1} J_{2n-1}(4 \beta_m) \sin \Biggl[ 2k \pi \frac{2n-1}{N} \Biggl] \label{bessel5} \\
              & = & 2 \cos \Biggl( \frac{2k \pi}{N} \Biggr) \sum^{\infty}_{n=1} J_{2n-1}(4 \beta_m) \sin \Biggl[ \frac{4nk \pi}{N} \Biggr] - 2 \sin \Biggl( \frac{2k \pi}{N} \Biggr) \sum^{\infty}_{n=1} J_{2n-1}(4 \beta_m) \cos \Biggl[ \frac{4nk \pi }{N} \Biggr]\label{bessel6} \\
\cos(2 \phi_k) & = & J_0 (2 \beta_m) + 2 \sum^{\infty}_{n=1} J_{2n}(2 \beta_m) \cos \Biggl[ \frac{4kn \pi}{N} \Biggl] \label{bessel7} 
\end{eqnarray} 

\hspace{.5cm} The arguments into the summation symbols in equations \ref{bessel2}, \ref{bessel4} and \ref{bessel6} and \ref{bessel7} are digital modulation frequency $W_m$ multiple integer, where $W_m = \frac{2 \pi}{N}$, where $N=\frac{f_s}{f_m}$. These arguments are made equal a multiple integer of $\frac{\pi}{2}$ by decimation \cite{netto} to reduce the output ripple factor signals due to several generated harmonics by nonlinearity signal response during the phase modulation. The decination factor best choice is made equal to $\frac{N}{4}$. Then, the \ref{bessel2}, \ref{bessel4} and \ref{bessel5} become: 

\pagebreak

\begin{eqnarray}
\sin( \phi^{d}_{k}) & = & -2 \sin \biggl( \frac{k \pi}{2} \biggr)\sum^{\infty}_{n=1}J_{2n-1}( \beta_m) \cos(nk \pi) \label{besseld1} \\
\sin(2 \phi^{d}_{k}) & = & -2 \sin \biggl( \frac{k \pi}{2} \biggr)\sum^{\infty}_{n=1}J_{2n-1}(2 \beta_m)cos(nk \pi)  \label{besseld2} \\
\sin(4 \phi^{d}_{k}) & = & -2 \sin \biggl( \frac{k \pi}{2} \biggr)\sum^{\infty}_{n=1}J_{2n-1}(4 \beta_m) \cos(nk \pi)  \label{besseld3} \\
\cos(2 \phi^{d}_{k}) & = & J_{0}(2 \beta_m) + 2 \sum^{\infty}_{n=1}J_{2n}(2 \beta_m) \cos(nk \pi) \label{besseld4}
\end{eqnarray}

\hspace{.5cm}The equations \ref{besseld1}, \ref{besseld2}, \ref{besseld3}  and \ref{besseld4}, after some algebraic manipulations, can be simplified to the following equations (see reference \cite{gradshteyn}):

\begin{eqnarray}
\sin( \phi^{d}_{k}) & = & (j)^{k-1} \biggl[ \frac{1-(-1)^k}{2} \biggr] \sin(\beta_m) \label{besselsimp1} \\
\sin(2 \phi^{d}_{k}) & = & (j)^{k-1} \biggl[ \frac{1-(-1)^k}{2} \biggr] \sin(2 \beta_m) \label{besselsimp2} \\
\sin(4 \phi^{d}_{k}) & = & (j)^{k-1} \biggl[ \frac{1-(-1)^k}{2} \biggr] \sin(4 \beta_m) \label{besselsimp3} \\
\cos(2 \phi^{d}_{k}) & = &  \cos^{2}(\beta_m)+ (-1)^{k} \sin^{2}(\beta_m)\label{besselsimp4}
\end{eqnarray}

\hspace{.5cm}The $\sin(m \beta_m)$ and $\cos(m \beta_m)$ values (m integer) are ever constants. Making $\Delta P(\Omega)$ equal $P_{ccw}(\Omega)-P_{cw}(\Omega)$, and replacing the terms of $\phi(k)$ sines and cosines  of equations \ref{gama0} to \ref{gama3} by equations \ref{besselsimp1} to \ref{besselsimp4}, the even terms are eliminated, giving a signal nonreciprocity, i.e., the rotation sense information. The resulting equation is given by:

\begin{equation}
\Delta P(\Omega)=2 \lbrack \Gamma^{1}_{k}(\pi) \Delta L'_{R}(\Omega) \sigma_0) + \Gamma^{3}_{k}(\pi) \Delta L'_{R}(\Omega) \sigma_0)^{3} + {\cal{O}}(5) \rbrack
\label{sinal1}
\end{equation}

\hspace{.5cm}The measured value of $\Omega$ is recursively computed using the \rm priori \rm values (at the $k-1$ instant). Then, we have the \rm priori \rm and \rm posteriori \rm expressions, resulting  the following expressions:

\begin{eqnarray}
\Delta P(\Omega) & = & 2 \lbrack \Gamma^{1}_{k}(\pi) \Delta L'_{R}(\Omega_{k}) \sigma_0) + \Gamma^{3}_{k}(\pi) \Delta L'_{R}(\Omega_{k}) \sigma_0)^{3} + {\cal{O}}(5) \rbrack
\label{sinal2} \\
\Delta P(\Omega) & = & 2 \lbrack \Gamma^{1}_{k}(\pi) \Delta L'_{R}(\Omega_{k-1}) \sigma_0) + \Gamma^{3}_{k-1}(\pi) \Delta L'_{R}(\Omega_{k-1}) \sigma_0)^{3} + {\cal{O}}(5) \rbrack
\label{sinal3}
\end{eqnarray}

\hspace{.5cm} Changing to the matricial form and neglecting the higher order and considering that the data aquisition time lack is very short to consider the among of $\Omega$ variation (the processing time is shorter than the $\Omega$ variation), the equations \ref{sinal2} and \ref{sinal3} can be put in the matricial form

\begin{equation}
\Biggl[  
							 \begin{array}{cc}
                                           \Gamma^{1}_{k} &  \Gamma^{3}_{k} \\
			                         \Gamma^{1}_{k-1} &  \Gamma^{3}_{k-1}
			                         \end{array} \Biggl]
 \Biggl[
							 \begin{array}{c}
                                           \pi \Delta L'(\Omega_k) \sigma_0  \\
			                         ( \pi \Delta L'(\Omega_k) \sigma_0 )^{3}			                        
							 \end{array}
\Biggl]
\approx \frac{1}{2} \Biggl[
							 \begin{array}{c}
                                           \Delta P(\Omega_k) \\
			                         \Delta P(\Omega_{k-1})			                        
							 \end{array}
\Biggl]
\label{matriz1}
\end{equation}

\hspace{.5cm}To computer the measured value of $\Omega$, after some matrix manipulations, the equation \ref{matriz1} can be written in the following equation

\begin{equation}
\frac{\Biggl[  
							 \begin{array}{cc}
                                           \Gamma^{1}_{k} &  \Gamma^{3}_{k} \\
			                         \Gamma^{1}_{k-1} &  \Gamma^{3}_{k-1}
			                         \end{array} \Biggl]^T
\Biggl[
							 \begin{array}{c}
                                           \Delta P(\Omega_k) \\
			                         \Delta P(\Omega_{k-1})			                        
							 \end{array}
\Biggl]
}{2 \lbrack \Gamma^{1}_{k} \Gamma^{3}_{k-1} - \Gamma^{1}_{k-1} \Gamma^{3}_{k} \rbrack} 
\approx \Biggl[
							 \begin{array}{c}
                                           \pi \Delta L'(\Omega_k) \sigma_0  \\
			                         ( \pi \Delta L'(\Omega_k) \sigma_0 )^{3}			                        
							 \end{array}
\Biggl]
\label{matriz2}
\end{equation}
\vspace{.5cm}

\hspace{.5cm}The modulator factor choice is very important because the unsuitable $ \beta_m $ value can cause a great overshoot, turning the system critically damped. In this work, the $ \beta_m $ is determinated by simulation. To correct the modulation factor, is needed to adjust the phase modulation driving amplifier gain. Note that the output matrix provides proportional to $ \Omega_k $ and $ \Omega^{3}_{k} $ signals, where the $ \Omega_k $ quantity is sent to the system output. 

\section*{\large \bf 3. The Setup and Operation Principle}
\vspace{.5cm}

\hspace{.5cm}The operation principle is based on syncronization field intensity combination, where the counterpropagating fields intensity are digitally processed. These signals are sent to a processing block separatelly into the time domain by the optical switch and electronic switch. Then, these signals are combined by difference between them, cancelling the even powers of $ \Omega_R $. The frequency decimation of these signal have the finallity of eliminate the modulation frequency harmonics that simplifies the processing alghoritm. The system setup is depicted in figure 2, where : PD=photodiode, LD=laser diode, PM=phase modulator, OS=optical switch, OC=optical coupler, BS=beam splitter, FOC=fibre-optic coil, PC=polarisation controller, OI=Optical isolator.

\vspace{.3cm}
\hspace{.5cm}The correlation effect and the offset can be corrected by an additional photodiode elimination and the employing of an optical switch pair (shown in figure 3), working syncronized with two CMOS switches together by a clock pulse train (the optical switch operation principle is described in reference \cite{saleh}). These optical switches can be integrated in a $LiNbO_3$ substrate. Each optical switch allows each field pass in a time interval equal of $\frac{T_s}{2}$ to the photodiode. The electrical switch scheme and electrical signal conditioning setup is proposed in figure 4 and 5, respectively, where $\Delta I_k$ is the generated signal difference by counterpropagating fields.

\hspace{.5cm}These optical switches provides a syncronizated signal difference with two CMOS switches, that separate the $I_cw$ and $I_ccw$ from photodiode at several instants of $\frac{T_s}{2}$. Note that the $I_ccw$ is delayed by a time lack of $\frac{T_s}{2}$. It is necessary that the $I_cw$ and $I_ccw$ signals arrive at same instant to the subtractor to avoid the signal nulling in equation \ref{sinal1}. These signals is subtacted one of each and decimated in frequency, explaned already in equations \ref{besseld1} to \ref{besselsimp4}. The difference signal is processed by a processing algorithm, which matricial equation is described in equation \ref{matriz2}. The modulating signal is used to generate the calculation matrix coefficients $\Gamma^{n}_{k}$ after frequency decimation. The initial values is set up to avoid overshoot. For small initial values, the overshoot reaches high values, otherwise, for high initial values, the setting time becomes very long and the system turn very slow. The ideal values can be foun by simulation. The best values of modulation factor $\beta_m$ minimizes the scale factor error and serves to maintain the modulation range limit into the FSR \footnote{FSR: \em Free Spectral Range \em (see reference \cite{saleh})} interval. The output can be sent to an adaptive LMS filter algorithm (see reference \cite{diniz}) to enhance the readout signal or at the digitalized input signal, or a predictive Kalman filter (an LMS adaptive filter scheme for the enhancement signal configuration is depicted in figure 6).

\section*{\large \bf 4. Simulation Results}

\hspace{.5cm}The simulations were made over a FORG wich uses a 10 meter lenght single mode, polarisation maintaining, fibre-optic, and 1.446 refraction index and $1.55 \mu m $ source lenght. The loop has 10 centimeter diameter, and the phase modulator is working at the modulation frequency equal 5.186722 MHz over the modulation factor $ \beta_m \approx 1.7278759594 $ . The sampling frequency is 100 times greather than the modulation frequency. These parameters  yields a scale factor error is around of $1.18 \times 10^{-3} \%$ at 20 radians per second is found. All these values are used in this work. The Signal response was plotted at 20rad/sec rotation rate within 200 samples (see figure 7), where $I_{ccw}(k)$ counterclockwise signal output, $I_{cw}(k)$ clockwise signal output, and signal difference.  These signals had been measured before pass by decimation frequency process.

In Figure 8, we can see the signal response at 20rad/sec rotation rate within 200 samples, where $I_{ccw}(k)$ counterclockwise signal output, $I_{cw}(k)$ clockwise signal output, and signal difference.  These signals had been measured after pass by zero order holding and decimation frequency process, and in the figure 9, we have the signal response at 20rad/sec rotation rate, the signal output at 200 samples, and the signal's FFT and phase response.

Reverting the rotation sense, note a signal response invertion (at -20rad/sec rotation rate:  signal output at 200 samples, FFT and phase response). Note the phase inversion in relation of the previous figure. At -2rad/sec (depicted in figure 11),  we get a signal response (showing signal output, FFT and phase response at 200 samples only), we can note that the FFT response varies proportionally (compare with the previous figure). It shows that the alghoritm is working properly and the equations shown in section 2 may appear valid.

\hspace{.5cm} But we can note the residual ripple due to the higher order expansion error approximation too (see figure 12). These riple can be easily minimized by low pass digital filtering. This fluctuation introduces a coloured noise, causing a bias drift at a mean value, masking the signal.

\vspace{.5cm}

\section*{\large \bf 5. Conclusions}

\hspace{.5cm}The use of an optical switching and digital frequency decimation can simplify the the optical and electronic hardware, reducing the photodiode offset, getting a signal directly proportional to the rotation rate with sense information. It can do the digital spectrometry processing method as an useful way to improve  a low cost resonant gyroscope performance using less fiber. In this technique, the FORG performance is dependent of modulation factor $ \beta_m $, as well as the decimation factor M, that smoot the response curve and reduces the processor's calculations cost.

\section*{\large \bf 6. Acknowledgements}

\hspace{.5cm}Thanks to God to realization of this work and for all, giving me energy and inspiration to conclude it. Thank to my wife Deise to her comprehension and her useful aids at each moment and incentivation in this work.

\end{document}